# PERSPECTIVE: THE CREATION OF NEWSGAMES AS A TEACHING METHOD - EMPIRICAL OBSERVATIONS

**Damien Djaouti,** LIRDEF - Montpellier University

**Julian Alvarez,** CIREL - University of Lille 1 / Play Research Lab - CCI Nord

## INTRODUCTION

The name serious game indicates "games that do not have entertainment, enjoyment or fun as their primary purpose" [1]. These games, which are being developed in increasing numbers, have useful applications in various market segments, including education, health, communication, politics, and defense. These different applications give rise to an organization of serious games into subcategories [2]. "Newsgames" are serious games that address current events or, as defined by [3]: "[G]ames that utilize the medium with the intention of participating in the public debate." While professional journalists, of course, use these games, sometimes individuals also use Newsgames to express their own points of view on current events [4].

As part of an introductory video game design course intended for engineering students, we opted to move away from the design of games for pure entertainment and focus on designing Newsgames. Although conducted empirically, introducing Newsgames into our course seemed to enrich the teaching. In addition to teaching a method for video game creation, these classes introduced other elements to our students as well. In this article, we will first present how our course is organized, then discuss the Newsgames created by our students and what they learned from this experience.

## EDUCATIONAL APPROACH

Since 2005, we have given a course on video game design to students with different majors at different training centers. The educational goal has been to encourage learners to design and execute small video game projects with a message. Games like these are known as serious games. For example, serious game themes focus on teaching about proper nutrition or introducing different sectors of industry. In 2010, we began focusing on designing Newsgames with engineering students majoring in computer science, civil engineering, chemical engineering, civil aviation, and meteorology. For three consecutive years, the classes were given at two different establishments, each with its own approach. The first establishment was a French engineering school called the INSA of Toulouse. This course was given as a one-week (35-hour) introductory module to students in their fourth year of their engineering master's. The second establishment was the computer science department of the French Toulouse III University. This course was only open to students in their fifth year of a computer science master's. A total of 30 hours of class time divided into two-hour sessions was spread out over one semester. The courses were structured in five phases:

1. **Introduction:** The learners basically had not heard of serious games at this juncture. We began by presenting concrete examples of games to captivate the learners and pique their curiosity. *Darfur is Dying* (2004) addresses the humanitarian crisis situation in Darfur and *September 12th* (2003) broaches the subject of responses to terrorism. These were the games that elicited many student questions: Who produces this game? Why did they design a game on this theme? Answers to these questions enabled us to introduce the serious game concept while presenting the various phases of the course.

2. **Discovery of serious games:** We then invited students to perform their own web search to identify other examples of serious games. The objective of this phase was for the learners to observe the wide variety of themes discussed through video games. This approach also inspired learners to design their own serious games.

3. **Video game design methods and tools:** We then introduced students to the fact that they would need to discuss a current event of their choosing through a video game, thereby designing a Newsgame-type of serious game. To do this, we held a class on entertainment video game design methods and tools adapted to serious games. We based this on our own design methodology, which is derived from the DICE model [5]. This generic design model, built by analyzing and summarizing a dozen different serious game design methodologies, outlines four major steps:

    o   Defining the serious content of the game.

    o   Imagining a game concept.

    o   Constructing a prototype.

    o   Evaluating the efficacy of the prototype.

    The last three steps form an iterative cycle that repeats until the designer is satisfied with the assessment of the serious game.

4. **Serious games design and execution:** The students divided themselves into groups of four to five people and began working on their serious game project. Throughout this phase, which represented the majority of the course in terms of hours, the teacher abandoned his "magisterial" role to adopt the role of a mentor accompanying students during the execution of their project.

5. **Presentation and evaluation of executed projects:** Finally, for evaluation purposes, the students presented their serious games to the rest of the class. A student who was not a member of the group was designated to test the game in front of everyone. This student was invited to give his or her opinion of the game, thereby eliciting discussions within the class. These presentations sometimes gave rise to real, in-depth debates among the students. We will see this in the next section.

To ensure that they successfully completed their serious game projects, our students were offered the tools they needed to do so. The engineering students attending our classes had different levels of computer expertise. There were two main groups. The first group included students familiar with a programming language, generally Java or C++. These were all engineering students majoring in computer science. The second group included engineering students from other areas. They usually had no special programming knowledge. In the latter group, which dominated our classes, we offered simple-to-use entertainment video game design software, like *RPG Maker* or *The Games Factory 2*. We call this type of application "game creation toolkit" [6] because they incorporate all the functions needed to design video games. In general, a two-hour game creation toolkit presentation session was enough to enable learners to start using the toolkit to create their own serious game.

**RESULTS: STUDENT PROJECTS**

From 2010 to 2012, we had approximately 80 engineering students in our Newsgames software design courses. These students designed a total of 17 games. The current events discussed and the software used to design these games are described in the following table.

**Table 1: Details of the Newsgames Designed by Students: Current Events Addressed and Design Software Used**

| Year | Current Event Addressed | Design Software Used |
|---|---|---|
| 2012 | The explosion in the number of millenarian sects predicting the end of the world in 2012. | The Games Factory 2 |
| 2012 | The Costa Concordia shipwreck (two games). | The Games Factory 2 |
| 2012 | The Megaupload closing (two games). | The Games Factory 2 (Group #1) <br> Java (Group #2) |
| 2012 | Scandals in the British press. | The Games Factory 2 |
| 2012 | The potential stripping of France's AAA credit status. | The Games Factory 2 |
| 2011 | Sharp fuel cost increases. | The Games Factory 2 |
| 2011 | The Mediator legal proceedings (Mediator is a French medication which lead to many unexpected deaths). | The Games Factory 2 |
| 2011 | The Wikileaks saga. | The Games Factory 2 |
| 2011 | The Arab Spring in Tunisia. | The Games Factory 2 |
| 2010 | The HADOPI law (a French law against Internet piracy). | Java |
| 2010 | The Haitian earthquake. | RPG Maker XP |
| 2010 | The Influenza A epidemic (two games). | Flash CS4 (Group #1) <br> Java (Group #2) |
| 2010 | The surge in suicides at Orange/France Telecom. | Flash CS4 |
| 2010 | The wave of expulsion of illegal immigrants by the French government. | Flash CS4 |

As illustrated in the table, the current events cover many themes. The students were able to choose the current event they wanted to develop. Using journalistic resources available on the web, the students found the documents they needed on their current event of choice. Subsequently, designing a video game based on news often encourages students to have in-depth debates on the current event in question.

For example, in 2010, the influenza A epidemic was the topic of a Newsgame called *Superflu*. The group of students who designed this game thought that the only way to eradicate the pandemic was through worldwide cooperation despite the differences in financial resources from country to country. The serious game of this group of students was a multiplayer game. Each player was assigned a part of the world with cities and vaccine or antiviral production sites. The players could choose to supply their own cities with the medications produced or supply cities managed by other players. The game was knowingly designed so that the only way to eradicate the pandemic was for all players to distribute all of the medications they produced to the first outbreak sites, even if these sites were outside of the player's own territory. This serious game conveyed the vision of one group of students in the class.

However, not every group had the same approach to a given current event. Another group from the same class designed *Flucorp Inc*. This group adopted a radically different view of the problem. Here, players represented directors of pharmaceutical production companies. The goal was to sell different products (protective equipment, vaccines, antivirals) with efficacy that varied according to how much was invested in medical research. If players invested massively in research, they could rapidly help contain the epidemic, but would not reap significant financial profits. In contrast, if players chose to invest only modestly in research, they would have the time to sell several different vaccines of ever-increasing efficacy. Subsequently, players were encouraged to manage the epidemic while maximizing profits. These students employed a nearly militant approach to the current event. Their game indirectly criticized the attitude adopted by pharmaceutical companies during this crisis. In the end, the different ways of addressing the same current event using Newsgames raised a lively debate between the two groups of students.

Some Newsgames designed by our students broached current events that were so sensitive they sometimes upset the players. The *Tunisian Oppression* game was about the 2011 Tunisian Revolution from the point of view of dictator Ben-Ali. Several small action games chronicled the epic of the dictator overthrown by his people. First, by playing the dictator's guards, players needed to contain the masses by shooting real bullets at them. The game was designed so that the people will always win. Players then needed to try to escape with as much gold as possible to a country willing to provide refuge. Despite its satirical undertones, this game sometimes offended. When students presented their game to the rest of the class, an energetic debate ensued on the way the game portrayed these events. Although the designers of this serious game wanted to express that "the people will always win despite the fury of the dictator," the discussions raised demonstrated that there is a certain difference in interpretation in the final message by the users.

This phenomenon occurred again with the *Escape from Port-au-Prince* game, the subject of which was the especially sensitive, horrific earthquake that hit Haiti in January 2010. Players represented catastrophe survivors who needed to leave the city and stay alive. The group of students who designed this serious game voluntarily decided to be fairly blunt regarding this current event. During their struggle to survive, players ran into pits filled with cadavers and were faced with Cornelian dilemmas to obtain food and conserve it. All this occurred while players were under the constant pressure of hunger and the risk of dying from it. When this project was presented at the end of our class, several students were literally shocked. In addition to the catastrophe itself, students questioned the way in which the topic was addressed."

Although they were video game lovers, and had themselves designed Newsgames, some students found this subject to be too serious to be addressed in a video game. The fact that violence was real and existed for people in the world, in contrast with what is found in entertaining video games bothered these students. In contrast, other students defended the validity of video games as media for addressing any subject, just like any other media. As a result, a spontaneous debate lasting about 30 minutes followed the presentation of this serious game.

In addition to everything they learned about video game design and the current events discussed, all students in this class reflected upon the possibilities and limitations of video games as a means of expression. This phenomenon is reminiscent of the *Darfur is Dying* serious game, which elicited similar reactions when we presented it to students at the beginning of the course. They were often shocked to play children who could be kidnapped by militia as they sought water in the desert during a war ongoing in real-life.

**CONCLUSION**

Given the various projects carried out by our student groups, we observed that designing Newsgames seems to stimulate debate on issues that go beyond the simple scope of video game design techniques. On the one hand, learners reinforced their knowledge about the current events they incorporated into their games thanks to the research they conducted, mainly via the Internet. On the other hand, it seems that by imposing "serious" subjects for a video game project, these learners exercised and developed their reasoning skills. For the purposes herein, a learner's "reasoning skills" means precisely documenting a subject to develop a point of view, and then comparing, contrasting, and defending this point of view vis-à-vis the point of view of other people. To this end, we identified three dimensions that seem to be interesting from a pedagogical point of view:

- **Development of an approach for addressing a given current event with support from journalistic information sources.** Designing a Newsgame based on a real subject requires learners to document the subject and acquire knowledge that can be used both to design the game and to elicit discussions during debate. Here, we observed that designing a video game, and a serious game in particular, encourages learners to read. This reading is generally thorough because the learners must be able to use the information they read in their video game.

- **Exchange of points of view on a given subject using video games as a medium for "discussion."** The fact that learner groups design Newsgames on a given subject using very different approaches shows them that not everyone sees current events in the same way. The example of the two influenza A games exemplifies two very different, fairly representative views of this subject.

- **Organization of student debates on the validity of video games as media for discussing current events.** We only observed this kind of debate for video games that incorporate very sensitive current events, such as the 2011 Tunisian Revolution or the 2010 Haitian earthquake. Although designing different video games for the same current event encouraged debate among our learners, the discussions moved away from the current event and moved toward the communication potential of video games. Therefore, in an informal way, there were discussions on issues related to media education.

Nevertheless, we observed that different learners do not subsume these three academic dimensions in the same way. Although all students took part in the first dimension (motivation to perform documentary searches), the same was not true for the second and third dimensions.

Only the groups of learners in a given class designing Newsgames on a given subject took part in debate on the current event incorporated into the video game. To promote debate on the communication potential of video games, we simply needed to identify at least one Newsgame that elicited a certain discomfort among a few learners in a given class.

To offer learners in a given class the opportunity to incorporate all three of these identified academic dimensions, perhaps it would be wise to require current event subjects to be sensitive and have different groups address the same subject in parallel.

According to our empirical observations, such an approach could certainly help develop the learners' reasoning skills. However, we can also question the potential of the educational scope of such an approach. For example, if we asked learners to create serious games other than Newsgames, such as learning games or edugames, would we observe the learners assimilating knowledge coming from an academic program?

When we refer to previous experiments conducted on the use of serious game design in classrooms [7-9], we do think that this teaching method has a real pedagogical potential that is yet to be fully explored. We will pursue our experiments with other students and other kinds of games, and we hope that this article will inspire you to try this method with your own students!